\documentstyle[prd,aps,preprint]{revtex}
\tightenlines
\newcommand{\gsim}{~\mbox{\raisebox{-1.0ex}{$\stackrel{\textstyle >}
{\textstyle \sim}$ }}}
\newcommand{\lsim}{~\mbox{\raisebox{-1.0ex}{$\stackrel{\textstyle <}
{\textstyle \sim}$ }}}

\newcommand{\bfk}{{\bf k}}

\newcommand{\order}{{\cal O}}
\newcommand{\beq}{\begin{equation}}
\newcommand{\eeq}{\end{equation}}
\newcommand{\beqa}{\begin{eqnarray}}
\newcommand{\eeqa}{\end{eqnarray}}

\newcommand{\lmk}{\left(}
\newcommand{\rmk}{\right)}
\newcommand{\lkk}{\left[}
\newcommand{\rkk}{\right]}

\begin{document}

\preprint{YITP-99-68, NI99021-SFU, OU-TAP-105, astro-ph/9911119}

\title{What happens when the inflaton stops during inflation}
\author{Osamu Seto$^1$, Jun'ichi Yokoyama$^{2,3}$, and Hideo Kodama$^{1}$}
\address{ $^1$Yukawa Institute for Theoretical Physics,\\ Kyoto
University, Kyoto 606-8502, Japan \\
$^2$Isaac Newton Institute for Mathematical Sciences,\\ University of
Cambridge,
Cambridge, CB3 0EH, UK \\
$^3$Department
of Earth and Space Science, Graduate School of Science,\\ Osaka
University, Toyonaka 560-0043, Japan }
\date{\today}
\maketitle

\begin{abstract}
The spectrum of adiabatic density perturbation generated during inflation
is studied in the case the time derivative of an inflation-driving scalar
field (inflaton) vanishes at some time during inflation.  It is shown
that the nondecaying mode of perturbation has a finite value even in this
case and that its amplitude is given by the standard formula with the
time derivation of the scalar field replaced by the potential gradient 
using the slow-roll equation.
\end{abstract}

\section{Introduction}

It is now widely believed that the large-scale homogeneity and
isotropy observed in the Universe were realized as a result of
accelerated expansion or inflation in the early universe \cite{inf}.
It also provides a mechanism to account for the origin of primordial
density fluctuations out of quantum fluctuations of the
inflation-driving scalar field which we call the {\it inflaton}
\cite{HS}.  In the standard inflation models such as new \cite{ni} and
chaotic \cite{ci} inflation, inflation is driven by the potential
energy of the inflaton as it slowly rolls the potential hill and is
predicted to produce adiabatic fluctuations with a nearly
scale-invariant spectrum.

More specifically, the amplitude of curvature perturbation, $\Phi$, on
comoving scale $r=2\pi/k$ is given by the formula,
\beq
   \Phi(r) \approx \left.\frac{H^2}{|\dot{\phi}|}\right|_{t_k},
\label{curv}
\eeq
where $\phi$ is the inflaton and $H$ is the Hubble parameter during
inflation, and the right-hand-side should be evaluated when the
relevant scale left the Hubble radius during inflation.  The above
formula also gives an estimate of the amplitude of density
perturbation on the comoving scale $r$ when this scale reentered the
Hubble horizon after inflation as well as that of large-scale
anisotropy of CMB due to the Sachs-Wolfe effect \cite{SW}, which has
been probed by the COsmic Background Explorer (COBE) satellite
\cite{COBE}.  The reason why (\ref{curv}) gives an almost
scale-invariant spectrum is that both $H$ and $\dot{\phi}$ change very
slowly during slow-roll inflation.

Recently, however, new classes of inflation models have been proposed
such that $\phi$ is not necessarily slowly rolling during the entire
period of inflation and that it changes direction of motion during
inflation.  One example is the oscillating inflation proposed by
Damour and Mukhanov \cite{oi,oiyuragi} in which accelerated expansion
is realized as the inflaton oscillates around a minimum of a nonconvex
potential.  In this model $\dot{\phi}$ vanishes twice in each period
of oscillation during inflation.  Another example is the chaotic new
inflation model proposed by one of us \cite{JY}.  This model assumes a
potential with a local maximum at the origin like new inflation but
start with the same initial condition as chaotic inflation.  If model
parameters are appropriately chosen, the inflaton climbs up the
potential hill near to the origin after chaotic inflation and new
inflation can be realized there.  In the early stage of new inflation
$\dot{\phi}$ may vanish and $\phi$ changes its direction of motion if
it does not have sufficient energy to go over the
origin.

In both models, if we apply the formula (\ref{curv}) as it is, the
amplitude of fluctuation apparently diverges when $\dot{\phi}$
vanishes.  The above formula, however, has been derived under the
slow-roll approximation, namely, under the assumption that both $\phi$
and $\Phi$ changes slowly during inflation.  On the other hand, motion
of $\phi$ is not given by the slow-roll formula when $\phi$ changes
its direction, because $\dot{\phi}$ vanishes where the gradient of the
potential does not vanish.

Thus we expect that it is inappropriate to apply (\ref{curv}) to the
case $\dot{\phi}$ vanishes during inflation.  The purpose of the
present paper is to derive a formula of curvature perturbation in such
a situation.  This is accomplished by a proper account of not only the
growing (or nondecreasing) mode but also the decaying mode as seen
below.  The rest of the paper is organized as follows.  In \S II we
give the formulation with an appropriate choice of a variable and
present solutions in the long- and the short-wavelength limit.  Then
these solutions are matched for the case of the linear potential in \S
III and for the quadratic potential in \S IV.  Section V is devoted to
the conclusion.

\section{Equations of motion and the solutions in the long- and the
short-wavelength limit }

We consider a minimally coupled singlet scaler field $\phi$ in
the spatially flat Robertson-Walker background metric,
\begin{equation}
ds^2 = -dt^2+a^2(t)d\textbf{x}^2,
\end{equation}
where $a(t)$ denotes the scale factor. Then the total action is
\begin{equation}
S = \frac{1}{2\kappa^2}\int R\sqrt{-g}d^4x+\int
\left[-\frac{1}{2}(\partial \phi)^2-V(\phi)\right]\sqrt{-g}d^4x,
\end{equation}
where $R$ is the scalar curvature and $\kappa^2=8\pi G$ with
 $G$ being the gravitational constant.
The Einstein equation and the field equation of   $\phi$ read
\beq
H^2 = \left(\frac{\dot{a}}{a}\right)^2=
\frac{\kappa^2}{3}\left(\frac{1}{2}\dot{\phi}^2+V(\phi)\right),\label{hubble}
\eeq \beq
\ddot{\phi}+3H\dot{\phi}+V'(\phi)=0. \label{eom.phi}
\eeq
with $H$ being the Hubble parameter.
Here a dot denotes time differentiation and a prime
represents differentiation with respect to $\phi$. These
are the equations for the unperturbed variables.

Next we incorporate a linear perturbation, writing the perturbed metric
in terms of the gauge-invariant variables  \cite{Baa} in the longitudinal
gauge,
\begin{equation}
ds^2
= -\lkk 1+2\Psi(\textbf{x},t)\rkk dt^2
+a^2(t)\lkk 1+2\Phi(\textbf{x},t)\rkk d\textbf{x}^2,
\end{equation}
where we use the notation of~\cite{KS} for the perturbation variables.
Hereafter all perturbation variables represent Fourier expansion
coefficients like
\begin{equation}
\Phi_\bfk=\int
\frac{d^3\textbf{x}}{(2\pi)^{3/2}}\Phi(\textbf{x},t)e^{i\textbf{kx}},
\end{equation}
and we omit the wave-number suffix $\bfk$.
We use the following combination of  gauge-invariant variables
~\cite{Mu,KH,NT},
\begin{equation}
Y=X-\frac{\dot{\phi}}{H}\Phi,
\label{defY}
\end{equation}
where
\begin{equation}
X=\delta\phi-\frac{a}{k}\dot{\phi}\sigma_g,
\end{equation}
is the gauge-invariant scalar field fluctuation
with $\sigma_g$ being the shear of each constant time
slice. The latter vanishes on Newtonian slice including the longitudinal
gauge.

The above quantity $Y$ is related to the gauge-invariant variable
\begin{equation}
  Z\equiv \Phi-\frac{aH}{k}V=\Phi
  + \frac{2}{3}\frac{\Phi+H^{-1}\dot{\Phi}}{1+w},
\end{equation}
as
\begin{equation}
Z=-\frac{H}{\dot{\phi}}Y,
\end{equation}
in the present case where matter consists of a scalar field.
Here $V$ is a gauge-invariant velocity perturbation \cite{KS} and $w$
denotes the ratio of pressure to energy density.
Although the definition of $Z$ is slightly different from the
Bardeen's $\zeta$ \cite{Ba}, which was originally defined by
\begin{equation}
\zeta = \lkk 1+\frac{2}{9}\frac{k^2}{(1+w)H^2a^2}\rkk \Phi-\frac{aH}{k}V,
\end{equation}
it enjoys the same property as $\zeta$, that is, both quantities are
constant during
$k \ll aH$ if only adiabatic fluctuation is present and sound velocity is
nonsingular.

From the perturbed Einstein equations, we obtain the following equations:
\begin{eqnarray}
\dot{X}\dot{\phi}-\ddot{\phi}X+\dot{\phi}^2\Phi=\frac{2}{\kappa^2}\frac{k^2}
{a^2}\Phi,
\\
\dot{\Phi}+H\Phi=-\frac{\kappa^2}{2}\dot{\phi}X,
\end{eqnarray}
From these equations and (\ref{defY}), the equation of motion of $Y$ reads
\begin{equation}
\ddot{Y}+3H\dot{Y}+\left[\left(\frac{k}{a}\right)^2+ M_{Y \rm eff}^2
  \right]Y=0,
\label{eom.Y}
\end{equation}
with
\beq
  M_{Y \rm eff}^2 \equiv V''(\phi)+3\kappa^2\dot{\phi}^
2-\frac{\kappa^4}{2H^2}\dot{\phi}^4+2\kappa^2\frac{\dot{\phi}}{H}V'(\phi).
\label{MYeff:def}\eeq
This equation has the following exact solution in the long-wavelength
limit $k \longrightarrow 0$ ~\cite{KH,H}.
\begin{eqnarray}
Y(t) &=& \tilde{c}_1(k)Y_1(t)+\tilde{c}_2(k)Y_2(t), \label{exsol} \\
Y_1(t) &=& \frac{\dot{\phi}}{H}, \\
Y_2(t) &=& \frac{\dot{\phi}}{H}\int^t_T\frac{H^2}{a^3\dot{\phi}^2}dt,
\end{eqnarray}
where $\tilde{c}_1(k)$ and $\tilde{c}_2(k)$
are integration constants to be determined by
quantum fluctuations generated during inflation, and $T$ is some
initial time which may be chosen arbitrarily
because its effect can be absorbed by a
redefinition of $\tilde{c}_1(k)$.
The solution $Y_2(t)$ is apparently singular at
$\dot{\phi}=0$. But in fact it is regular
there \cite{KH}.

Since $Y_2(t)$ has a different dimension than $Y_1(t)$, 
it is more convenient to
redefine the coefficients as
\begin{equation}
   c_1(k)\equiv \tilde{c}_1(k),~~~~~c_2(k)\equiv \tilde{c}_2(k)/k^3,
\end{equation}
for which the scale factor appears in the rescaling-invariant form of
$k/a(t)$, and $c_1(k)$ and $c_2(k)$ have the same
dimension.

We next consider the evolution of $Y$ in short-wavelength regime 
in order to set the initial condition of $Y$ out of quantum fluctuations.
Since we are interested in generation of perturbations around the turning 
point of the inflaton, we only consider the evolution of $Y$ in the inflationary stage around and after the time $\dot \phi=0$ and assume 
that $|\dot H/H^2|\ll 1$ holds in this stage. We further assume that  
$\mu \equiv V'/(\kappa V)$ satisfies the condition 
\begin{equation}
\mu'\le 0, \quad \mu \mu'' \ge 0,
\label{V:cond}\end{equation}
for the value of $\phi$ in this stage. This condition is satisfied for 
quite a large class of potentials including the pure exponential potential and those which are approximately given by $c|\phi|^n$ or $V_0-c|\phi|^n$ around the turning point where $c>0$ and $n$ is a positive integer.

$M_{Y\rm eff}^2$ defined by (\ref{MYeff:def}) can be rewritten in 
terms of $\mu$ as%
\begin{equation}
\frac{M_{Y\rm eff}^2}{\kappa^2 V}=\frac{1}{\kappa}\mu'
+ \left(\frac{\kappa\dot \phi}{H}+\mu\right)^2.
\label{MYeffBymu}
\end{equation}
As is shown in the appendix, under the condition (\ref{V:cond}), the 
value of $|\mu'|/\kappa$ at the turning point should be smaller than 
$1/N$, where $N$ is the number of $e$-folds of inflation after the turning point, 
{\it i.e.}, the increase of $\ln a$. There it is also shown that 
$|\kappa\dot\phi/(H\mu)|$ is always less than unity and 
approaches unity, {\it i.e.}, $\kappa\dot\phi/(H\mu) \rightarrow -1$, in a 
few expansion time after $\phi$ passes the turning point. 
This implies that the second term in the right-hand side of 
(\ref{MYeffBymu}) is equal or less than the value of $2|\dot H/H^2|$ 
at the slow-roll phase following the short transient period around 
the turning point. To be precise, this term is approximately given 
by $(2/9)(|\mu'|/\kappa)^2|\dot H/H^2|$ in the slow-roll phase.

From these estimates we see that $M_{Y\rm eff}^2$ in (\ref{eom.Y}) 
can be neglected in the regime
\begin{equation}
\left(\frac{k}{aH}\right)^2 > \frac{1}{N},
\end{equation}
which extends to a wavelength much larger than the Hubble horizon 
size. Thus in this regime (\ref{eom.Y}) becomes
\begin{equation}
\ddot{Y}+3H\dot{Y}+\left(\frac{k}{a}\right)^2Y = 0.
\end{equation}
Under the condition $|\dot H/H^2|\ll1$ the solution of this equation 
satisfying the normalization condition for positive frequency modes 
to define the vacuum state,
\begin{equation}
a^3(Y\dot{Y}^* -\dot Y Y^*) = i,
\end{equation}
is approximately given by
\begin{equation}
Y=\frac{iH}{\sqrt{2k^3}}\left[\alpha_\bfk (1+ik\eta)e^{-ik\eta}
-\beta_\bfk (1-ik\eta)e^{ik\eta}\right],
\end{equation}
where $\eta=-1/(Ha)$, and $\alpha_\bfk$ and $\beta_\bfk$ are 
constants which satisfy
\begin{equation}
|\alpha_\bfk|^2-|\beta_\bfk|^2 = 1.
\end{equation}
We shall choose $(\alpha_\bfk \,,\beta_\bfk ) = (1,0)$ so that the 
vacuum reduces to the one in Minkowski spacetime at the 
short-wavelength limit ($-k\eta \rightarrow\infty$). Therefore we 
obtain
\begin{equation}
Y=\frac{iH}{\sqrt{2k^3}}(1+ik\eta)e^{-ik\eta}.
\label{ksol}
\end{equation}

\section{Matching the short- and the long-wavelength regimes: The
case of a linear potential}

Having obtained the short- and the long-wavelength solutions we next
connect these two regimes in order to determine the coefficients
$c_1(k)$ and $c_2(k)$.  In this section we consider the simplest case
in which the inflaton's potential can be approximated by a linear function
near the point where the inflaton switches its direction of motion.
That is, we take,
\begin{equation}
V(\phi)=V_0+V'\phi,
\label{pot}
\end{equation}
with $V_0$ and $ V'$ being  constants and assume $|V'\phi| \ll V_0$.
Then the solution of (\ref{eom.phi}) is
\begin{equation}
\dot{\phi}(t)=\dot{\phi}_s\lmk 1-e^{-3H(t-t_0)}\rmk
=\dot{\phi}_s\lkk 1-\lmk\frac{a(t_0)}{a(t)}\rmk^3\rkk,
\label{phidot}
\end{equation}
where $\dot{\phi}_s\equiv -V'/3H$ is the velocity of
$\phi$ in the slow-roll limit and we have set
\begin{eqnarray}
\dot{\phi}(t_0)=0.
\end{eqnarray}
In this case (\ref{exsol}) is rewritten as
\begin{equation}
Y(t)=c_1(k)\frac{\dot{\phi}_s}{H}\lkk 1- \lmk\frac{a(t_0)}{a(t)}\rmk^3
\rkk +c_2(k)\frac{-
1}{3\dot{\phi}_s}\lmk\frac{k}{a(t)}\rmk^3,
\label{sol1}
\end{equation}
where we have chosen $T\longrightarrow \infty$
 so that $Y_2(t)$ becomes pure decaying
mode. It should be noted that it is nontrivial that $Y_2(t)$ is described
like
this all the time because the integrand  is singular
at $t=t_0$ ~\cite{KH}. We  emphasize that $Y_1(t)$, which is the
 nondecreasing mode, has
a decaying component in contrast to the case of slow-roll inflation.

We next match (\ref{sol1}) with the short-wavelength solution (\ref{ksol})
via two different methods and confirm both methods give the same results.

The first method we adopt is to use the series expansion of (\ref{ksol})
with respect to $k/a(t)$, namely,
\beq
  Y =
\frac{iH}{\sqrt{2k^3}}
\lkk 1+\frac{1}{2}\lmk\frac{k}{a(t)H}\rmk^2
+\frac{i}{3}\lmk\frac{k}{a(t)H}\rmk^3+
\ldots\rkk, \label{eksol}
\eeq
and match each coefficient of the power series with that of the
long-wavelength solution.
As is seen here the zeroth-order term matches the main part of
the nondecreasing term and the third-order term corresponds to the
decaying mode.  However, the second-order term, which is present in
(\ref{eksol}), is absent in (\ref{sol1}).
In fact, this term should be regarded as a finite-wavenumber correction
to the nondecreasing mode $Y_1$ as we see now.

With the help of the $k=0$ solutions $Y_1(t)$ and $Y_2(t)$ and
the Green function method, we obtain from (\ref{eom.Y})
the following iterative
expression for arbitrary $k$ \cite{KH}.
\begin{equation}
Y = c_1(k)Y_1+c_2(k)k^3Y_2+k^2Y_1\int aY_2Ydt-k^2Y_2\int aY_1Ydt.
\end{equation}
The lowest-order iterative solution valid for small $k$ reads
\begin{equation}
Y(t) \cong
c_1(k)Y_1(t)+c_2(k)k^3Y_2(t)
+c_1(k)\lmk\frac{k}{a(t)}\rmk^2\frac{\dot{\phi}_s}{2H^3}
+\order\lmk\lmk \frac{k}{a(t)}\rmk^5\rmk . \label{ite}
\end{equation}
The new term corresponds to the second term in (\ref{eksol})
and we can now determine $c_1(k)$ and $c_2(k)$ consistently
up to the third order in $(k/a)^3$, to yield,
\begin{eqnarray}
c_1(k) &=& \frac{iH}{\sqrt{2k^3}}\frac{H}{\dot{\phi}_s}, \label{c1}\\
c_2(k)
&=&\frac{\dot{\phi}_s}{\sqrt{2k^3}H^2}\lkk
1-3i\lmk\frac{a(t_0)H}{k}\rmk^3\rkk.
\label{c2}
\end{eqnarray}
We note that $a(t_0)H/k=1$ for the mode leaving the Hubble radius when
$\dot{\phi}$ vanishes.

Another method to decide $c_1(k)$ is simply to equate the value and
the first time derivative of (\ref{ksol}) and (\ref{ite}) at
the watershed.
Because  $M_{Y \rm eff}^2$ is $\order(H^2/N)$,
we can use the
short-wavelength  solution during $(k/a)^2 \gsim H^2/N$ and the
long-wavelength solution for $(k/a)^2 \lsim H^2/N$.
Thus we define the dividing epoch
$\eta_c$ by $(k/a(\eta_c))^2=H^2/N$ and match
(\ref{ksol}) and (\ref{ite}) there.

At $\eta=\eta_c$ the short-wave solution (\ref{ksol}) and its derivative
read
\beqa
   Y(\eta_c)&=&\frac{iH}{\sqrt{2k^3}}\lkk 1+ \order\lmk(-k\eta_c)^2\rmk\rkk,
\label{ys}\\
   \frac{dY}{d\eta} (\eta_c)&=&\frac{iH}{\sqrt{2k^3}}k^2\eta_c\lkk
1-ik\eta_c +
   \order\lmk(-k\eta_c)^2\rmk\rkk . \label{yds}
\eeqa
On the other hand, from (\ref{ite}) the long-wave counterparts are given by
\beqa
   Y(\eta_c)&=&c_1(k)\frac{\dot{\phi}_s}{H}\lkk 1+ \frac{1}{2}(-k\eta_c)^2
  -\lmk\frac{\eta}{\eta_0}\rmk^3\rkk -
c_2(k)\frac{H^3}{3\dot{\phi}_s}(-k\eta_c)^3,
  \label{yl}\\
   \frac{dY}{d\eta}(\eta_c)&=&c_1(k)\frac{\dot{\phi}_s}{H}\lmk k^2\eta_c
  - 3\frac{\eta_c^2}{\eta_0^3}\rmk +
c_2(k)\frac{H^3}{\dot{\phi}_s}k^3\eta_c^2.
  \label{ydl}
\eeqa
Equating (\ref{ys}) with (\ref{yl}) and (\ref{yds}) with (\ref{ydl})
we find
\beqa
c_1(k) &=& \frac{iH}{\sqrt{2k^3}}\frac{H}{\dot{\phi}_s}
\lkk 1+ \order\lmk(-k\eta_c)^2\rmk\rkk, \label{c1a}\\
c_2(k)
&=&\frac{\dot{\phi}_s}{\sqrt{2k^3}H^2}\lkk 1-3i\lmk\frac{a(t_0)H}{k}\rmk^3
+\order\lmk -k\eta_c\rmk \rkk,
\label{c2a}
\eeqa
in agreement with (\ref{c1}) and (\ref{c2}).  Note that the leading terms of
the
above results are independent of the choice of the matching epoch $\eta_c$.
Thus we find
\beq
  c_1(k)=-\frac{3iH^3}{\sqrt{2k^3}V'}.
\eeq

\if
 At $\eta_=\eta_c$, we
obtain by (\ref{ksol}) and (\ref{s})
\begin{eqnarray}
Y(\eta_c) & \sim & \frac{iH}{\sqrt{2k^3}} \label{1} \\
\frac{dY}{d\eta}(\eta_c) & \sim
& -\frac{iH}{\sqrt{2k^3}}k\sqrt{\frac{|\dot{H}|}{H^2}}
\end{eqnarray}
and by (\ref{sol1}) and (\ref{s})
\begin{eqnarray}
Y(\eta_c) & = &
c_1\frac{\dot{\phi}_s}{H}[1-(\eta_c/\eta_0)^3]+c_2\frac{H^3}{3\dot{\phi}_s}\
eta_c{}^3 \nonumber \\
 & \sim & c_1\frac{\dot{\phi}_s}{H}+c_2\frac{H^3}{3\dot{\phi}_s}\eta_c{}^3
\label{3} \\
\frac{dY}{d\eta}(\eta_c) & =
& -c_1\frac{\dot{\phi}_s}{H}3\eta_0{}^{-3}\eta_c{}^2+c_2\frac{H^3}{\dot{\phi
}_s}\eta_c{}^2 \label{4}
\end{eqnarray}
The validity of approximation to (\ref{3}) is due to
\begin{eqnarray}
(\eta_c/\eta_0)^3 &=&
\frac{1}{(-k\eta_0)^3}\left(\frac{|\dot{H}|}{H^2}\right)^3 \\
 & \ll & 1
\label{neg}
\end{eqnarray}
with help of (\ref{s}) and $(-k\eta_0)>\order(1)$ because now we
focus attention on the wavenumbers which has horizon crossing time
$t_k$ such that $t_k/t_0\geq\order(1)$.

From (\ref{1})-(\ref{4}),(\ref{s}) and (\ref{neg}), we obtain
\begin{eqnarray}
c_1\frac{\dot{\phi}_s}{H} \sim \frac{iH}{\sqrt{2k^3}} \\
c_2\frac{H^3}{3\dot{\phi}_s}\eta_c{}^3 \ll c_1\frac{\dot{\phi}_s}{H}
\end{eqnarray}
Therefore we see
\begin{equation}
c_1(k) = \frac{iH}{\sqrt{2k^3}}\frac{H}{\dot{\phi}_s}
\end{equation}
\fi

\section{The case of a quadratic potential}

Next we consider the case for which the potential is locally given by,
\begin{equation}
V(\phi)=V_0-\frac{1}{2}m^2\phi^2,
\end{equation}
near the turning point, where $V_0$ and $ m$ are constants satisfying
$H^2\equiv 8\pi GV_0/3 \gg m^2$.  We assume $V_0 \gg m^2\phi^2$.
Then the solution of (\ref{eom.phi}) reads
\begin{eqnarray}
\dot{\phi}(t) =
\frac{\lambda_+}{1-\frac{\lambda_+}{\lambda_-}}\phi_0\left(e^{\lambda_+(t-t_
0)}-e^{\lambda_-(t-t_0)}\right), \\
\lambda_\pm \equiv -\frac{3}{2}H\pm\sqrt{\frac{9}{4}H^2+m^2},
\end{eqnarray}
where we have set
\begin{eqnarray}
\dot{\phi}(t_0) &=& 0, \\
\phi(t_0) &=& \phi_0.
\end{eqnarray}
By virtue of the inequality $H \gg m$ we find
\beq
  \lambda_+ \cong \frac{m^2}{3H^2},~~~~~\lambda_- \cong -3H.
\eeq
To the lowest order in $m^2/(3H^2)$, (\ref{exsol}) is expressed as
\beq
  Y=c_1(k)\frac{m^2}{3H^2}\phi_0\lkk 1-\lmk\frac{a(t_0)}{a(t)}\rmk^3\rkk
  -c_2(k)\frac{H}{m^2\phi_0}\lmk\frac{k}{a(t)}\rmk^3. \label{ms}
\eeq
We can repeat the same argument as the case of a linear potential.  As is suggested by
the
fact that (\ref{ms}) coincides with (\ref{sol1}) by the replacement,
\[
   -\frac{m^2\phi_0}{3H}=-\frac{V'[\phi_0]}{3H} \Longleftrightarrow
   \dot{\phi}_s=-\frac{V'}{3H},
\]
we obtain in this case,
\beq
   c_1(k)=-\frac{3iH^3}{\sqrt{2k^3}V'[\phi_0]}.
\eeq
Thus in both cases the coefficient of the nondecaying mode is given by
the standard formula as if $\phi$ is slowly rolling with the speed,
\beq
    \dot{\phi}=-\frac{V'[\phi]}{3H},
\eeq
even for the $k-$mode leaving the horizon when $\dot{\phi}$ vanishes.

\if
Therefore (\ref{exsol}) is rewritten as
\begin{eqnarray}
Y(t)&=&c_1(k)\frac{1}{H}\frac{\lambda_+}{1-\frac{\lambda_+}{\lambda_-}}\phi_
0\left(e^{\lambda_+(t-t_0)}-e^{\lambda_-(t-t_0)}\right)  \nonumber \\
& &
+c_2(k)\frac{-1}{3}\left(\frac{1-\frac{\lambda_+}{\lambda_-}}{\lambda_+\phi_
0}\right)\frac{e^{-\frac{3}{2}H(1-\sqrt{1+(2m/3H)^2})t_0}}{\sqrt{1+(2m/3H)^2
}}e^{-\frac{3}{2}H(1+\sqrt{1+(2m/3H)^2})t}
\label{sol2}
\end{eqnarray}
As above-mentioned, the condition
\begin{equation}
\left(\frac{2m}{3H}\right)^2 \ll 1
\end{equation}
is satisfied during inflation. In this limit, (\ref{sol2}) coincides
with (\ref{sol1}) and there is this correspondence of
\begin{equation}
\dot{\phi}_s = \frac{-V'}{3H} \Leftrightarrow \lambda_+\phi_0 =
\frac{-V'(\phi_0)}{3H}
\label{cor}
\end{equation}

The correlativity (\ref{cor}) suggests that the transformation rule
\begin{equation}
\frac{-V'}{3H} \rightarrow \frac{-V'(\phi_0)}{3H}
\end{equation}
would be useful for general potentials.
\fi

\section{Conclusion}

In the present paper we have derived a formula for the spectrum of
density perturbation generated by inflation in the case the
time-derivative of the inflaton, $\dot \phi$, vanishes at some time
during inflation.  Calculation in such a situation is not
straightforward because the amplitude of growing mode in the
long-wavelength limit vanishes at the turning point of the inflaton,
which implies that the Bardeen parameter estimated for quantum
fluctuations at horizon crossing diverges at that time. This is the
origin of the divergence of the standard formula at $\dot\phi=0$. Of
course, the amplitude of perturbations generated around the turning
point do not diverge in reality because the Bardeen parameter becomes
finite when the inflaton moves away from the turning point. This
implies that we must subtract a contribution of decaying mode, for
which the Bardeen parameter is not constant, from the
perturbation generated from quantum fluctuation in order to obtain a
correct formula for the amplitude of growing mode.

In order to treat this problem properly, we have first shown that the
standard expression for quantum fluctuations in the short-wavelength
limit gives a good approximation for the evolution of perturbations
during inflation, at least in the case we are concerned with, even
when wavelengths are much larger than the Hubble horizon size. Then
we have matched this expression with an exact long-wavelength
expression for the evolution of the perturbation variable, $Y$, which
does not show any singular behavior when $\dot{\phi}$ vanishes, in a
range of wavelength in which both expressions are valid to determine
the amplitude of growing mode in quantum fluctuations.  As a
result we have obtained the new formula for the amplitude of curvature
perturbation
\begin{equation}
\Phi\left(r=\frac{2\pi}{k}\right) =
f\left.\frac{3H^3}{2\pi |V'[\phi]|}\right|_{t_k},
\end{equation}
where $f=3/5~ (2/3)$ in the matter- (radiation-) dominated stage.
Apparently this formula coincides with the standard formula with
$\dot{\phi}$ in the denominator replaced by the slow-roll velocity,
\begin{equation}
  \dot{\phi}=-\frac{V'(\phi)}{3H},  \label{slow}
\end{equation}
but it is valid not only in the slow-roll regime but also in an extreme
situation $\dot{\phi}=0$ when the relevant mode leaves Hubble radius during
inflation.

Although we have done calculations only for two simple models in the
present paper, the above formula seems to hold for perturbations
generated around and after the turning point of the inflaton in
generic models for the following reason. First recall that the
expression (\ref{ksol}) for perturbations generated from quantum fluctuations
gives a good approximation during inflation even when the wavelength
is much larger than the Hubble horizon size. Though 
it is a combination of a growing mode and a decaying mode at the horizon 
crossing, the amplitude of the latter component decreases rapidly
with cosmic expansion after the horizon crossing. In
that stage the Bardeen parameter $Z=-HY/\dot\phi$ should approach a
constant. As shown in the appendix, in a several Hubble time
after passing the turning point, the slow-roll approximation becomes
good and $\dot\phi/H$ is expressed as  $-V'/3H^2$, which is constant 
with a good accuracy in the early phase of inflation after the turning 
point. In the meanwhile $Y$ given by (\ref{ksol}) also becomes a constant
$iH/\sqrt{2k^3}$. Hence the value of the Bardeen parameter for the
growing mode should be given by
\begin{equation}
Z=-i\frac{H}{\sqrt{2k^3}}\frac{3H^2}{V'}.
\end{equation}
This expression coincides with $Z$ obtained from $c_1(k)Y_1$ with
$c_1(k)$ given by (\ref{c1}). This argument does not determine the
amplitude of the decaying-mode component, but it has no importance in
the estimation of curvature perturbation in later states.

\acknowledgements
{
One of the authors (JY) would like to thank Isaac Newton Institute for
Mathematical
Sciences for its hospitality where this research was completed.
This work was supported in part by the Monbusho Grant-in-Aid for Scientific
Research
Nos.\ 11740146 (JY) and 11640273 (HK), and ``Priority Area: Supersymmetry and
Unified Theory
of Elementary Particles (No.\ 707)'' (JY).}

\appendix
\section*{}

In this appendix, for inflation models in which $\dot \phi$ vanishes 
at some time during inflation and the inflaton potential $V$ 
satisfies the condition (\ref{V:cond}), we show that 
$|(V'/(\kappa^2V))'|$ is much smaller than unity around the turning 
point of the inflaton as well as in the slow-roll phase of the 
inflationary stage after that. We also show that a local quadratic 
approximation for the potential $V$ gives a good description of the 
evolution of the inflaton and the scale factor in the same period.

First we rewrite the basic evolution equations (\ref{hubble}) and
(\ref{eom.phi}) in terms of the new variables
\begin{eqnarray}
&&v =\kappa\frac{\dot{\phi}}{H},\\
&&\zeta=\ln (a/a_0)
\end{eqnarray}
as 
\begin{eqnarray}
&& \kappa\frac{d\phi}{d\zeta}=v,\label{v:eq1}\\
&& 
\frac{dv}{d\zeta}=-3\left(1-\frac{v^2}{6}\right)\left(v+\mu\right), 
\label{v:eq2}\end{eqnarray}
where $a_0$ is the value of $a$ at the time when $\dot\phi=0$. $H$ is
written in terms of these variables as
\begin{equation}
H^2=\frac{\kappa^2 V}{3\left(1-\frac{v^2}{6}\right)}.
\end{equation}
Note that from the equation
\begin{equation}
\frac{\dot H}{H^2}=-\frac{1}{2}v^2,
\end{equation}
the inflation condition $(aH)\dot{}>0$ is expressed as $v^2<2$.

For definiteness we consider the case $\mu>0$ at the turning point 
$\phi=\phi_0, v=0$. In this case $v$ becomes negative after the 
inflaton passes the turning point. We restrict the consideration to 
the inflationary stage $0\le \zeta \le \zeta_*$, where $\zeta_*$ is 
the value of $\zeta$ at the time when $v$ becomes $-\sqrt{2}$ for 
the first time. In this stage (\ref{v:eq2}) yields
\begin{equation}
\frac{d}{d\zeta}\left(1+\frac{v}{\mu}\right)
=-3\left[1-\frac{v^2}{6}+\frac{\nu}{3}\left(1+\frac{|v|}{\mu}\right)
\right]\left(1+\frac{v}{\mu}\right)+\nu,
\label{v/mu:ODE}\end{equation}
where $\nu=-\mu'/\kappa\ge0$. From this equation we immediately see 
that $1+v/\mu$ cannot vanish. Hence (\ref{v:eq2}) is written as%
\begin{equation}
\frac{d|v|}{d\zeta}=3\left(1-\frac{v^2}{6}\right)(\mu -|v|).
\label{v:eq2'}\end{equation}

From the condition $\nu'\le0$, $\mu$ satisfies 
$\mu'/\kappa\le-\nu_0$, where $\nu_0$ is the value of $\nu$ at 
$\phi=\phi_0$. Integrating this equation, we obtain
\begin{equation}
\mu \ge \mu_0 +\nu_0\kappa(\phi_0-\phi),
\end{equation}
where $\mu_0=\mu(\phi_0)$. From this equation and 
(\ref{v:eq2'}) it follows that $\phi$ satisfies the differential 
inequality
\begin{equation}
\frac{d^2\phi}{d\zeta^2}+2\frac{d\phi}{d\zeta}
+\frac{2\mu_0}{\kappa}+2\nu_0(\phi_0-\phi)
\equiv f\le0.
\end{equation}
Solving this equation yields
\begin{eqnarray}
&|v(\zeta)|=& \frac{\mu_0}{(1+2\nu_0)^{1/2}}
\left(e^{\lambda_1\zeta}-e^{-\lambda_2\zeta}\right)\nonumber\\
&&-\frac{\lambda_1}{2(1+2\nu_0)^{1/2}}e^{\lambda_1\zeta}
\int_0 d\zeta e^{-\lambda_1\zeta}f
-\frac{\lambda_2}{2(1+2\nu_0)^{1/2}}e^{-\lambda_2\zeta}
\int_0 d\zeta e^{\lambda_2\zeta}f,
\end{eqnarray}
where 
\begin{equation}
\lambda_1=\sqrt{1+2\nu_0}-1,\quad
\lambda_2=\sqrt{1+2\nu_0}+1.
\end{equation}
Since $f\le0$, this equation gives the inequality
\begin{equation}
|v|\ge \frac{\mu_0}{(1+2\nu_0)^{1/2}}
\left(e^{\lambda_1\zeta}-e^{-\lambda_2\zeta}\right).
\end{equation}
Hence we obtain the estimate 
\begin{equation}
N\equiv \zeta_* \le \frac{1}{2\lambda_1}\ln \frac{2(1+2\nu_0)}{\mu_0}
\sim \frac{1}{\nu_0}.
\end{equation}
In particular it follows that $\nu_0$ should be much smaller than 
unity if a sufficient inflation occurs after the inflaton passes the 
turning point.

Next we show that $v/\mu$ rapidly approaches unity. Integrating 
(\ref{v/mu:ODE}), we obtain
\begin{equation}
1+\frac{v}{\mu}=e^{-3g(\zeta)}
+\int_0^\zeta d\zeta_1 
\nu(\zeta_1)e^{-3(g(\zeta)-g(\zeta_1))},\end{equation}
where 
\begin{equation}
g(\zeta)=\int_0^\zeta d\zeta_1 \left[1-\frac{v^2}{6}
+\frac{\nu}{3}\left(1+\frac{|v|}{\mu}\right)\right].
\end{equation}
Since $g$, $|v|$ and $\nu$ are nondecreasing functions of $\zeta$, 
this equation gives the inequality
\begin{equation}
1+\frac{v}{\mu}\le \frac{2\nu}{6-v^2+2\nu}
+\frac{6-v^2}{6-v^2+2\nu}
\exp\left[-\frac{1}{2}(6-v^2+2\nu_0)\zeta\right].
\end{equation}
Therefore we obtain 
\begin{eqnarray}
&&1+\frac{v}{\mu}\lsim \frac{\nu}{2},\label{1+v/mu:bound}\\
&&\left|\frac{1}{v}\frac{dv}{d\zeta}\right| \lsim \nu,
\end{eqnarray}
for $\zeta\gsim\zeta_s\equiv\frac{1}{3}\ln\frac{3}{\nu}$. 
Since $\frac{dv}{d\zeta}/v$ is expressed as
\begin{equation}
\frac{1}{v}\frac{dv}{d\zeta}
=\frac{\ddot\phi}{H\dot\phi}-\frac{\dot H}{H^2},
\end{equation}
the above estimate implies that the slow-roll approximation becomes 
good for $\zeta\gsim \zeta_s$, {\it i.e.}, in a few Hubble time 
after the inflaton passes the turning point.

Finally we show that the linear approximation $\mu\simeq \mu_0 
+\nu_0(\phi_0-\phi)$ for $\mu$ can be used to determine the 
evolution of the inflaton with a good accuracy for a long period 
after the inflaton enters the slow-roll regime. First from 
{(\ref{v:eq2'}) and (\ref{1+v/mu:bound}) $|v|$ is estimated as
\begin{equation}
|v|\lsim \left(1+\frac{3}{2}\nu\zeta\right)\mu.
\end{equation}
Integrating (\ref{v:eq1}) by taking account of this inequality, we 
obtain
\begin{equation}
\frac{\nu}{\mu}\kappa|\phi-\phi_0|\lsim 
\nu\zeta \left(1+\frac{3}{4}\nu\zeta\right).
\end{equation}
If we expand $\mu$ as 
\begin{equation}
\mu=\mu_0+\nu_0\kappa(\phi_0-\phi)+\cdots,
\end{equation}
the ratio of the second term and the subsequent higher-order terms 
is in general of the order of $(\nu_0/\mu_0)\kappa|\phi_0-\phi|$. 
Hence the above estimate implies that the first two terms in this 
expansion gives a good approximation for $\mu$ in the period $\zeta 
< 1/\nu_0$, which covers a large fraction of the slow-roll stage.

\end{document}